\documentstyle[12pt]{article}
\begin{document}

\tolerance=5000

\def\cL{{\cal L}}
\def\be{\begin{equation}}
\def\ee{\end{equation}}
\def\bea{\begin{eqnarray}}
\def\eea{\end{eqnarray}}
\def\tr{{\rm tr}\, }
\def\nn{\nonumber \\}
\def\gd{g^\dagger}
\def\e{{\rm e}}

\  \hfill 
\begin{minipage}{3.5cm}
NDA-FP-27 \\
April 1996 \\
hep-th/yymmxxx \\
\end{minipage}

\vfill

\begin{center}

{\Large\bf
Exact Four Dimensional Exploding Universe Solution
in String Theory
}

\vfill

{\Large\sc Shin'ichi NOJIRI}\footnote{
e-mail : nojiri@cc.nda.ac.jp}

\vfill

{\large\sl Department of Mathematics and Physics \\
National Defence Academy \\
Hashirimizu Yokosuka 239, JAPAN}

\vfill

{\bf ABSTRACT}
\end{center}

We obtain a four dimensional exploding 
universe solution in string theory. The 
solution is obtained from the string 
theory in the flat background by using 
non-abelian $T$-duality and the analytic 
continuation. In the solution, the 
radius of the universe is finite for 
fixed time and the universe is 
surrounded with the boundary which is 
composed of the singularity. The 
boundary runs away with the speed of 
light and the flat space-time is left 
behind.

\vfill


\newpage

It would be interesting to consider the cosmological 
problems in the string theories since the string 
theory is a theory of the gravity.
In this paper, we give a four dimensional 
exploding universe 
solution of the string theories. 
The radius of the universe is finite for fixed time and 
is surrounded by the boundary which is composed 
of singularity. 
The boundary runs away with the speed of light and 
the flat space-time is left behind.

The solution is obtained from the string 
theory in the flat background by using non-abelian 
$T$-duality \cite{quevedo} and the analytic continuation.
When a $\sigma$ model 
in two dimensions has an abelian or non-abelian isometry, 
we can introduce the gauge fields 
by gauging the isometry and impose a constraint which  
makes the gauge curvature vanish. 
After imposing a gauge fixing condition and integrating 
gauge fields, we obtain the dual model.
In this model, there appears a 
shift of the dilaton fields when we use a 
regularization which keeps the general covariance. 
The non-abelian $T$-duality can be used to show 
the equivalence of string models corresponding to 
different topologies \cite{alvarez}. 
Recently the equivalence of the chiral model and 
Wess-Zumino-Witten model \cite{WZW} 
when coupled with gravity was shown \cite{chiral} 
by using the model dual to the chiral model \cite{zachos}.

We start with four dimensional space with Euclidean 
signature by assuming that the extra dimensions 
(twenty-two dimensions in bosonic string 
or six dimensions in superstring) are 
properly compactified.
We parametrize the four coordinates 
$X^\mu$ ($\mu=0,1,2,3$) 
by 
\be
\label{coord}
g=\e^{\rho+\phi^i\sigma^i}=X^0+iX^i\sigma^i\ .
\ee
Here $\sigma^i$'s ($i=1,2,3$) are Pauli matrices.
Then the Lagrangean of the string in four dimensional 
flat background is given by
\bea
\label{flatlag}
\cL&=&{1 \over 2}\bar\partial X^\alpha \partial X^\alpha \nn
&=&{1 \over 4}\tr \bar\partial \gd \partial g \ .
\eea
The Lagrangean has $SO(4)\sim SU(2)_L \otimes SU(2)_R$ 
symmetry.
The two $SU(2)$ transformations are given by
\bea
&&g\rightarrow hg\ \ \ (SU(2)_L)\ ,\ \ \ \ 
g\rightarrow gh\ \ \ (SU(2)_R) \ .
\eea
($h$ is a group element of $SU(2)$.)
In order to obtain a dual Lagrangean, we gauge  
$SU(2)_L$ symmetry by introducing gauge fields $A$ 
and $\bar A$,
\bea
\label{gaugef}
&&A=A^iT^i\ ,\ \ \ \bar A=\bar A^iT^i \\
&&T^i={1 \over 2}\sigma^i\ , 
\eea
\be
\label{gauged}
\cL^{\rm gauged}=-{1 \over 4}\tr \gd (\partial g+ A g)
\gd (\bar\partial g + \bar A g) 
\ee
and add a term which makes the gauge curvature 
$F=[\partial + A, \bar\partial + \bar A]$ vanish,
\be
\label{constraint1}
\cL^{\rm constraint}={1 \over 2}\tr \xi F
\ee
Here $\xi$ is an element of $SU(2)$ algebra
\be
\label{constr}
\xi=\xi^i T^i\ .
\ee
When we integrate the gauge fields $A$ and $\bar A$ by 
choosing the gauge condition
\be
\label{condition}
\phi^i=0\ ,
\ee
we obtain the dual Lagrangean
\be
\label{dual}
\cL^{\rm dual}=
{1 \over 2}\e^{2\rho}\partial\rho\bar\partial\rho
+{1 \over \e^{4\rho} + 4 (\xi)^2}
\left[2\e^{2\rho}\delta_{ij}-4\epsilon_{ijk}\xi^k +8\e^{-2\rho}\xi^i\xi^j
\right]\partial\xi^i\bar\partial\xi^j \ .
\ee
Here $(\xi)^2=\xi^i\xi^i$. 
Furthermore, by using the regularization which keeps 
the general covariance, we find that there appears 
a dilaton term in the dual theory \cite{quevedo},
\bea
\label{dilaton}
\cL^{{\rm dilaton}}&=&
-{1 \over 4\pi}R^{(2)}\Phi \nn
\Phi&=&\ln (\e^{4\rho} + 4 (\xi)^2)\ .
\eea
Here $\Phi$ is the dilaton field.
Eq.(\ref{dual}) tells that the target space metric 
is given by
\be
\label{dualmet}
d{\bf s}^2=
{1 \over 2}\e^{2\rho}d\rho^2
+{1 \over \e^{4\rho} + 4 (\xi)^2}
\left[2\e^{2\rho}\delta_{ij}
+8\e^{-2\rho}\xi^i\xi^j
\right]d\xi^i d\xi^j \ .
\ee
When we express $\xi^i$'s by the polar coordinates 
$(r,\theta,\varphi)$
\be
\label{polar}
\left\{ \begin{array}{l}
\xi^1=r\sin\theta\cos\varphi \\
\xi^2=r\sin\theta\sin\varphi \\
\xi^3=r\cos\theta
\end{array}
\right.
\ee
we obtain
\be
\label{metric1}
d{\bf s}^2={1 \over l}(dl^2 + dr^2)
+{l r^2 \over l^2 + r^2}(d\theta^2 + \sin^2\theta d\varphi^2)\ .
\ee
Here
\be
\label{l}
l=\e^{2\rho}\ .
\ee
By further changing the coordinates $(l,r)$ to $(t,s)$ by
\bea
\label{s}
s^2&=&{l r^2 \over l^2 + r^2} \\
\label{t}
t^2&=&s^2 x^{-1}(2-3x)^{1 \over 6}(1-x)^{-{1 \over 2}}
\eea
we find
\bea
\label{metricE}
d{\bf s}^2&=&{4 \over 4 - 11x+ 8x^2}\left\{
(2-3x)^{11 \over 6}(1-x)^{-{1 \over 2}}dt^2 
+ ds^2 \right\} \nn
&&+s^2 (d\theta^2 + \sin^2\theta d\varphi^2)\ .
\eea
Here $x$ is defined by
\be
\label{x}
x\equiv {s^2 \over l}={r^2 \over l^2 + r^2}
\ee
and is a function of $(t,s)$ given by solving Eq.(\ref{t}).
The metric can be analytically continuated by 
\bea
\label{ancon}
&& s\longrightarrow is \\
&& \left( \begin{array}{l}
x\longrightarrow -x \\
t^2\longrightarrow t^2=
s^2 x^{-1}(2+3x)^{1 \over 6}(1+x)^{-{1 \over 2}}
\end{array} \right)
\eea
and becomes the metric of Lorentz signature
\bea
\label{metricL}
d{\bf s}^2&=&{4 \over 4 + 11x+ 8x^2}\left\{
(2+3x)^{11 \over 6}(1+x)^{-{1 \over 2}}dt^2 
- ds^2 \right\} \nn
&&- s^2 (d\theta^2 + \sin^2\theta d\varphi^2)\ .
\eea
As we will see in the following, the metric (\ref{metricL}) 
describes the exploding universe 
whose radius is finite for fixed $t$ and 
the universe is surrounded with the boundary
of the singularity, which runs away with the speed of 
light and the flat space-time is left behind.

When $s$ is fixed and $t$ goes to infinity or 
$s$ goes to zero and $t$ is fixed, the metric 
(\ref{metricL}) has the following form
\be
\label{szero}
d{\bf s}^2\sim 2^{11 \over 6}dt^2-ds^2
-s^2 (d\theta^2 + \sin^2\theta d\varphi^2)\ .
\ee
This tells that there is no conical singularity at $s=0$ 
and the metric approaches to the metric in the 
flat background when $t$ becomes large.

On the other hand, when $s$ is fixed and $t$ goes 
to zero or 
$s$ goes to infinity and $t$ is fixed, the metric 
(\ref{metricL}) has the following form
\be
\label{sinf}
d{\bf s}^2\sim 3^{7 \over 4}2^{-1}\left(s^{-1}tdt^2
-3^{-2}s^{-3}t^3ds^2 \right)
-s^2 (d\theta^2 + \sin^2\theta d\varphi^2)\ .
\ee
This tells that there is a singularity at $t=0$.
Furthermore, when $s$ goes to infinity and $t$ is fixed, 
the ratio of the length scale in the angular 
($\theta$ or $\phi$) direction to that in the radial 
($s$) direction diverges and there is a singularity at $s=\infty$. 
Note that the distance $R$ between the 
singularity ($s=\infty$) and 
the center ($s=0$) is finite and given by
\bea
\label{dising}
R&=&\int_0^\infty ds{2 \over \sqrt{4+11x+8x^2}} \nn
&=&t\int_0^\infty dx{(4+11x+8x^2)^{1 \over 2} \over 
x^{1 \over 2} (2+3x)^{13 \over 12} (1+x)^{3 \over 4}}\ .
\eea
$R$ can be regarded as the radius of the universe and 
is proportional to $t$.
In fact, the singularity at $s=\infty$ runs away with 
the speed of light.
In order to see this, we define Kruskal-like 
coordinates $\zeta^\pm$ by
\be
\label{Kr}
\zeta^\pm\equiv s^2 x^{-1} \left\{ 1 
\pm \left( {x \over 1+x} \right)^{1 \over 2}
 \right\}\ .
\ee
By using these coordinates $\zeta^\pm$, the 
metric (\ref{metricL}) can be rewritten by 
\be
\label{metricK}
d{\bf s}^2 = {2 \over \zeta^+ + \zeta^-}d\zeta^+d\zeta^- 
-s^2 (d\theta^2 + \sin^2\theta d\varphi^2)\ .
\ee
Here $s^2$ is given by
\be
\label{ssquare}
s^2={(\zeta^+ + \zeta^-)(\zeta^+ - \zeta^-)^2 
\over \zeta^+ \zeta^-}
\ee
In these coordinates, the orbit of radially radiated light 
is given by the line where $\zeta^+$ or $\zeta^-$ is a 
constant.
Eq.(\ref{ssquare}) tells that the singularity at 
$s^2=\infty$ corresponds to 
$\zeta^+=0$ or $\zeta^-=0$. 
Therefore the singularity, which can be 
regarded as a boundary of the universe, runs away with the 
speed of light.
Since the dilaton field $\Phi$ in Eq.(\ref{dilaton}) 
is given in terms of the coordinates $\zeta^\pm$ by 
\be
\label{dilatonK}
\Phi=\ln \zeta^+\zeta^-\ , 
\ee
the dilaton field $\Phi$ becomes singular if and only if 
$s^2=\infty$ ($\zeta^+=0$ or $\zeta^-=0$).

In summary, we have obtained a four dimensional 
exploding universe solution in string theory. 
The solution is obtained from the string 
theory in the flat background by using non-abelian 
$T$-duality and the analytic continuation. 
In the solution, the radius of the universe is finite 
for fixed time and the universe is surrounded with 
the boundary of the singularity.
The boundary runs away with the speed of light 
like the wave front of an explosion and 
the flat space-time is left behind.
Since the solution is obtained from the string theory 
of the flat background, we can exactly calculate the 
correlation functions on the world sheet.

\ 

\noindent
{\bf Acknowledgement}

The author would like to thank A. Sugamoto for discussions. 
He is also indebted to K. Fujii who called his attention to 
non-abelian $T$-duality.

\end{document}